\documentclass[aps,pra,twocolumn,groupedaddress,showpacs]{revtex4}

\usepackage[dvips]{graphicx}
\usepackage{indentfirst}

\def\be{\begin{equation}}
\def\ee{\end{equation}}
\def\bea{\begin{eqnarray}}
\def\eea{\end{eqnarray}}

\begin{document}

\title{Method for monopole creation in spinor Bose-Einstein condensates}

\author{D.E. Chang}
\email{darrick.chang@utu.fi}
\affiliation{Department of Physics, University of Turku, FIN-20014 Turun yliopisto, Finland}

\date{\today}

\begin{abstract}
We present a method for creating a monopole in an antiferromagnetic spin-1 Bose-Einstein condensate.  The required phase engineering of the multicomponent condensate is achieved using light shifts, which depend on both the magnetic substate $m_F$ and polarization of the incident laser beam.
\end{abstract}


\maketitle

%
\section{Introduction}
%
Since the experimental realization of Bose-Einstein condensates (BEC's) in 1995~\cite{anderson95,davis95,bradley95,bradley97}, a wide variety of properties and effects in BEC's has been studied.  Among the major topics of interest are topological defects such as vortices~\cite{baym96,marzlin97,dum98,matthews99,dobrek99,martikainen01} and solitons~\cite{dum98,reinhardt97,burger99,denschlag00}.  More recently, the realization of BEC's in optical traps~\cite{stamperkurn98,barrett01} has opened up interesting new avenues of research in this area.  This development makes it possible to create multicomponent condensates where all magnetic substates $m_F$ are trapped.  The resulting spin degree of freedom permits the existence of topological defects that cannot appear in scalar condensates, such as skyrmions~\cite{khawaja01} and monopoles~\cite{stoof01}.  Important properties of monopoles, such as stability and dynamics, have already been examined ~\cite{stoof01,martikainen02}.  Here we propose an explicit method for creating a monopole in BEC's, focusing on the experimentally relevant case of ${}^{23}\mathrm{Na}$.  Our proposal extends the method of ``phase imprinting'' used to generate topological defects in scalar condensates ~\cite{dobrek99,burger99,denschlag00}.  In particular, we demonstrate that the energy shift experienced by each component of the condensate depends on both $m_F$ and the polarization of the laser beam, and that by applying a sequence of different polarizations the phases of the components can be engineered in a broad manner.

A spin-1 BEC can be described by the condensate wavefunction ${\Psi}_i=\sqrt{n(\mathbf{r})}\zeta_{i}(\mathbf{r})$, where $n(\mathbf{r})$ is the total density and $\zeta(\mathbf{r})$ is a normalized spinor describing the local spin of the condensate.  For an anti-ferromagnetic BEC such as ${}^{23}_{}\textrm{Na}$, the ground state spinor is given by~\cite{ho98}
\be
\label{eq:ground}
\zeta_{g}(\mathbf{r})=\left(
\begin{array}{c}
0 \\
1 \\
0
\end{array}\right).
\ee

The spinor corresponding to a monopole has the form~\cite{stoof01}
\be\label{eq:monopolespinor}
\zeta_{m}(\mathbf{r}) = \frac{1}{\sqrt{2}}\left
(\begin{array}{c}
-m_{x}(\mathbf{r})+im_{y}(\mathbf{r}) \\
\sqrt{2}m_{z}(\mathbf{r}) \\
m_{x}(\mathbf{r})+im_{y}(\mathbf{r})
\end{array}\right).
\ee
$\mathbf{m}(\mathbf{r})$ is a unit vector field that describes the local rotation of the ground state spinor.  By considering the energy and winding number of such a condensate, it can be shown~\cite{stoof01} that $\mathbf{m}(\mathbf{r})=\mathbf{r}/r$ uniquely characterizes the monopole structure.  We see that $\mathbf{m}(\mathbf{r})$ assumes the familiar hedgehog structure associated with a monopole.

For the purpose of phase engineering, it is convenient to also write this spinor in spherical coordinates as
\be\label{eq:spherical}
\zeta_{m}(\mathbf{r}) = \frac{1}{\sqrt{2}}\left
(\begin{array}{c}
-\sin {\theta} e^{-i\phi} \\
\sqrt{2}\cos \theta \\
\sin {\theta} e^{i\phi} \\
\end{array}\right),
\ee
where $\phi$ is the azimuthal angle and $\theta$ the polar angle.  In addition, we define the phase of each component $\varphi_{i}\: {\equiv}\: \arg(\zeta_i)$.  For example, the phases of the components in the monopole are given by
\be
\label{eq:monopolephase}
\varphi_{m}=
\left(\begin{array}{c}
-\phi+\pi \\
\pi\theta(-z) \\
\phi
\end{array}\right),
\ee
where $\theta(z)$ is the step function.
%
\section{Condensate dynamics}
%
The Hamiltonian of the condensate in second quantized form is~\cite{ho98,ohmi98}
{\setlength\arraycolsep{2pt}
\begin{eqnarray}
\mathcal{H} & = & \int \mathrm{d}\mathbf{r} \left(\frac{{\hbar}^2}{2M} \nabla\psi_{i}^{\dagger}\cdot\nabla\psi_{i} + V_{\textrm{\small{trap}}}({\mathbf{r}})\psi_{i}^{\dagger}\psi_{i}+{}\right. \nonumber \\
& & \left. {} + \frac{c_0}{2}\psi_{i}^{\dagger}\psi_{j}^{\dagger}\psi_{j}\psi_{i} + \frac{c_2}{2}\psi_{i}^{\dagger}\psi_{j}^{\dagger}\mathbf{F}_{il}\cdot\mathbf{F}_{jk}\psi_{k}\psi_{l}\right) + {} \nonumber \\
& &
{} + \mathcal{H}_{B}+\mathcal{H}_{L},\label{eq:Hamiltonian}
\end{eqnarray}}
{\noindent}where $\psi_{i}(\mathbf{r})$ is the field annihilation operator for spin state $i$, $F_i\ \{i=x,y,z\}$ are the spin matrices, $V_{\textrm{\small{trap}}}({\mathbf{r}})$ is the trapping potential, and $c_0=4\pi{\hbar}^{2}(a_0+2a_2)/3m$ and $c_2=4\pi{\hbar}^{2}(a_2-a_0)/3m$ are constants that depend on the scattering lengths $a_0$ and $a_2$.  $\mathcal{H}_{B}$ and $\mathcal{H}_{L}$ describe respectively the applied magnetic field and the laser beams that will be used to create the monopole.  During the times that the applied magnetic field and lasers are turned on, we assume that their respective Hamiltonians dominate $\mathcal{H}$ and are the only terms that need to be considered in the evolution of the system.

The Hamiltonian for the applied magnetic field is given by $\mathcal{H}_{B}  = -\int d\mathbf{r}\, \psi_{i}^{\dagger}(\gamma\mathbf{B}\cdot\hbar\mathbf{F}_{ij})\psi_{j}$, where $\gamma$ is the gyromagnetic ratio.  When we ignore all other terms in $\mathcal{H}$, the magnetic field leads to the following equations of motion:
\be\label{eq:B-motion}
\frac{\mathrm{d}}{\mathrm{dt}}{\Psi_{i}} =  \frac{i}{\hbar}(\gamma\mathbf{B}\cdot\hbar\mathbf{F}_{ij})\Psi_{j}.
\ee

\begin{figure*}
\begin{center}
\includegraphics[width=7cm]{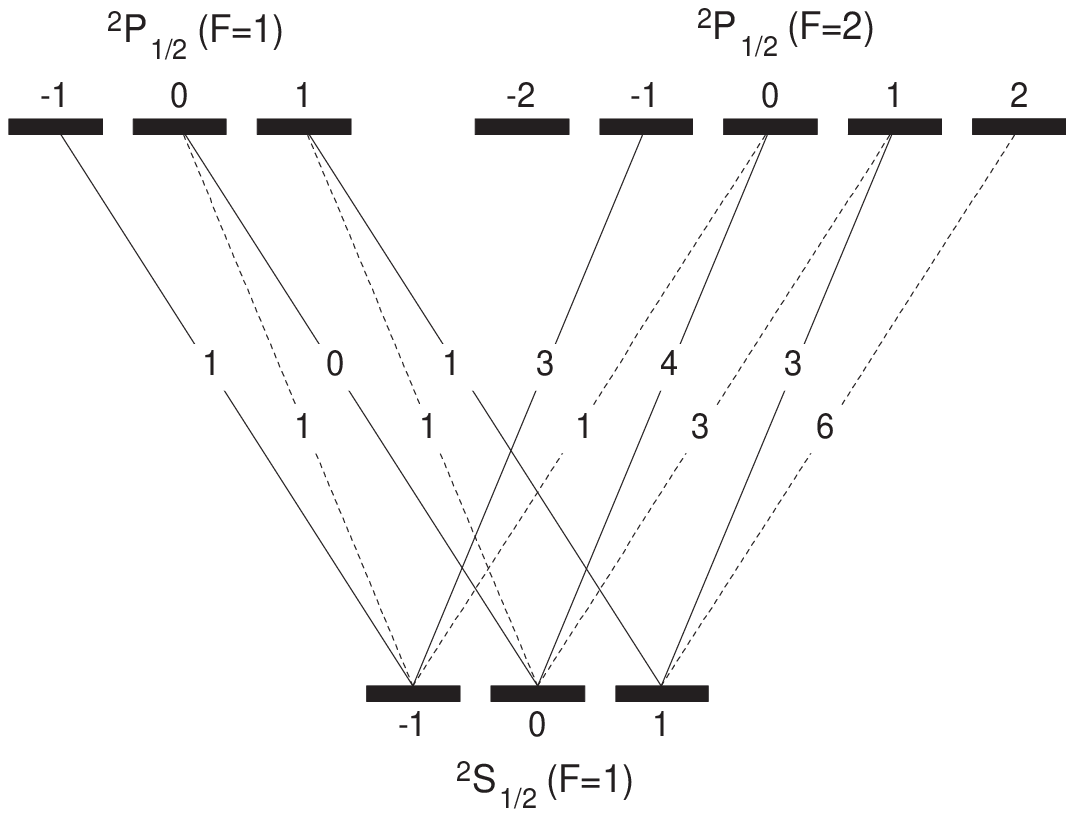}
\end{center}
\caption{The values for $C_{ij}^2$, scaled so that the smallest value is an integer.  The solid lines give the coefficients for $\pi$-polarized light, while the dashed lines correspond to $\sigma^+$-polarized light.  The values for $\sigma^-$-polarized light are obtained from the $\sigma^+$ values by making the substitutions $m_F{\rightarrow}-m_F$ in the diagram.\label{fig:transitionstrengths}}
\end{figure*}

When a laser beam is applied, the ground state energy of magnetic substate $i$ is shifted by an amount~\cite{grimm00}
\be
\label{eq:lightshift}
{\Delta}E_{i}=\frac{3{\pi}c^2\Gamma}{2\omega_0^3}I\times\sum_{j}\frac{C_{ij}^2}{\delta_{ij}},
\ee
where the sum is carried out over all excited states $|j\rangle$.  Here, $\delta_{ij}$ is the detuning of the laser from the $|i\rangle\rightarrow|j\rangle$ transition, $I$ the laser intensity, $\Gamma$ the rate of spontaneous decay, $\omega_0$ the atomic resonance frequency, and $C_{ij}$ is a transition strength characterizing the coupling between atom and laser.  The coefficients $C_{ij}$ depend on the ground substate $m_F$ and on the polarization of the laser beam and are given in Figure~\ref{fig:transitionstrengths}.
%
\section{Phase engineering using phase imprinting}
%
The concept of phase engineering via imprinting was suggested~\cite{dobrek99,denschlag00} and experimentally demonstrated~\cite{burger99,denschlag00} as a means of creating topological defects in scalar condensates.  We extend the method to spinor condensates, focusing on the case of ${}^{23}\mathrm{Na}$.  We show that the energy shifts induced by different beam polarizations allow for wide capabilities in phase engineering.

\begin{figure*}
\begin{center}
\includegraphics[width=4cm]{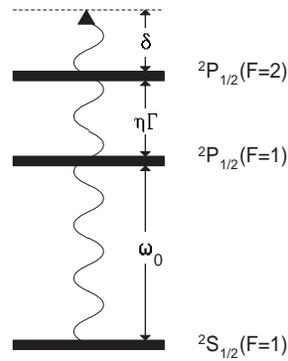}
\end{center}
\caption{The relevant frequencies of the atom-laser system.\label{fig:frequencydiagram}}
\end{figure*}

Suppose that a laser is detuned by an amount $\delta$ from the ${}^{2}S_{1/2}(F=1)\: {\rightarrow}\: {}^{2}P_{1/2}(F=2)$ transition, and that the states ${}^{2}P_{1/2}(F=1)$ and ${}^{2}P_{1/2}(F=2)$ are separated in frequency by an amount $\eta\Gamma$, as shown in Figure~\ref{fig:frequencydiagram}.  From Eq.~(\ref{eq:lightshift}), the energy of the ground state is shifted by an amount
\be
{\Delta}E_{i}=\frac{3{\pi}c^2\Gamma}{2\omega_0^3}I\times\sum_{{}^{j{\in}P_{1/2}(F=1)}_{k{\in}P_{1/2}(F=2)}}\left(\frac{C_{ij}^2}{\delta+\eta\Gamma}+\frac{C_{ik}^2}{\delta}\right).
\ee
We assume that the laser pulse has a square envelope with a duration $t$ that is short compared to the condensate's correlation time $t_c=\hbar/\mu$, where $\mu$ is the chemical potential.  In this case the primary effect of the light shift is to imprint a phase onto each component~\cite{burger99}:
\be
\label{eq:phaseshift}
\zeta_{i}(\mathbf{r})\rightarrow\zeta_{i}(\mathbf{r})\mathrm{exp}[-i{\Delta}E_{i}t/\hbar].
\ee

Using the coefficients given in Figure~\ref{fig:transitionstrengths} and the expansion $\frac{1}{\delta+\eta\Gamma}=\frac{1}{\delta}-\frac{\eta\Gamma}{\delta^2}+\cdots$, one can obtain explicit results for the phase shifts generated in the condensate.  For $\delta\gg\eta\Gamma$, the shifts corresponding to $\pi$-, $\sigma^{+}$-, and $\sigma^{-}$-polarized beams are
\begin{eqnarray}
\Delta\varphi^{\pi}(t)^{T} & \approx & -{\alpha}t[(2\quad 2\quad 2)-\frac{\eta\Gamma}{2\delta}(1\quad 0\quad 1)]\label{eq:pi-phaseshift}, \\
\Delta\varphi^{+}(t)^{T} & \approx & -{\alpha}t(3\quad 2\quad 1)\label{eq:plus-phaseshift}, \\
\Delta\varphi^{-}(t)^{T} & \approx & -{\alpha}t(1\quad 2\quad 3)\label{eq:minus-phaseshift}.
\end{eqnarray}
Here $\alpha\; {\equiv}\; {\pi}c^{2}{\Gamma}I/4\hbar{\omega_0^3}\delta$.  Equations~(\ref{eq:pi-phaseshift})-(\ref{eq:minus-phaseshift}) suggest a basic technique for phase engineering of the spinor condensate.  To shift the phase of $m_F=1$ relative to $m_F=-1$, one can use beams with polarizations $\sigma^+$ and $\sigma^-$.  Once the phase between these two states is fixed as desired, one can use $\pi$-polarized light to adjust the phase of $m_F=0$ relative to the other states.  Although it is not used in our proposal for monopole creation, we note that this phase shift is due to a second order term in Eq.~(\ref{eq:pi-phaseshift}), and we thus expect that it will be more difficult to implement in an experiment.  It may be useful, in addition, to alter the global phase of the condensate, which can be realized approximately by using the first order term in Eq.~(\ref{eq:pi-phaseshift}).  Finally, one can incorporate spatial dependence into the phases by using the above pulses in conjunction with absorption plates to create non-uniform intensity profiles $I(\mathbf{r})$~\cite{dobrek99}.  Writing $I(\mathbf{r})=I_{0}\rho(\mathbf{r})$, where $I_0$ is the maximum intensity in the profile and $0\leq\rho(\mathbf{r})\leq{1}$, Eqs.~(\ref{eq:pi-phaseshift})-(\ref{eq:minus-phaseshift}) remain valid with the replacements $\alpha\rightarrow\alpha_{0}\rho(\mathbf{r})$ and $\alpha_{0}\; \equiv\; {\pi}c^2{\Gamma}I_{0}/4\hbar\omega_{0}^{3}\delta$.
%
\section{Model for monopole creation}
%
Martikainen et al.~point out~\cite{martikainen02} that the monopole spinor given in Eq.~(\ref{eq:monopolespinor}) resembles a combination of other topological defects.  Indeed, the form of $\zeta_{m}(\mathbf{r})$ presented in Eq.~(\ref{eq:spherical}) makes it clear that the $m_F=+1$ state contains a vortex, the $m_F=-1$ state an anti-vortex, and the $m_F=0$ state a soliton.  They thus suggest that one can experimentally create monopoles by creating the appropriate defects in the individual components.  Here we provide a method to accomplish this using the phase imprinting concepts discussed above.

Starting from the ground state of Eq.~(\ref{eq:ground}), we first apply a constant magnetic field of magnitude $B_0$ along the negative $y$-axis for a time $t_{B}$.  Defining $\omega_{B}={\gamma}B_0$, Eq.~(\ref{eq:B-motion}) readily yields
\be
\zeta_{B}\equiv\zeta(t_{B})=\frac{1}{\sqrt{2}}
\left(\begin{array}{c}
-\sin(\omega_{B}t_{B}) \\
\sqrt{2}\cos(\omega_{B}t_{B}) \\
\sin(\omega_{B}t_{B})
\end{array}\right).
\ee
The phases of $\zeta_B$ (for $0<\omega_{B}t_{B}<\pi/2$) are
\be
\varphi_{B}^{T}=(\pi{\quad}0{\quad}0).
\ee
The effect of the magnetic field is to populate the $m_F={\pm}1$ states prior to phase engineering.  Experimentally, $t_B\ (0<\omega_{B}t_{B}<\pi/2)$ should be chosen to best realize the monopole structure.

To create the monopole, we assume that we have blue-detuned lasers, all of which have equal intensities $I_0$ and values of $\delta$, and which can be applied in square-shaped pulses.  The results can be accordingly modified to cover more general situations.  In addition, we assume that we have absorption plates that can create the following intensity profiles:
\begin{eqnarray}
\rho_{1}(\mathbf{r}) & = & \frac{\phi}{2\pi}\label{eq:+azimuthalintensity}, \\
\rho_{2}(\mathbf{r}) & = & \frac{(2{\pi}-\phi)}{2\pi}\label{eq:-azimuthalintensity}, \\
\rho_{3}(\mathbf{r}) & = & \theta(z)\label{eq:stepintensity}, \\
\rho_{4}(\mathbf{r}) & = & \theta(-z).
\end{eqnarray}
The pulse sequence to create the monopole is given in Table~\ref{table:pulsesequence}.  Pulses A and B use intensity profiles with azimuthal dependence to create a vortex in the $m_F=1$ component and an anti-vortex in the $m_F=-1$ component.  Pulses C and D together imprint the necessary phases to generate the soliton in the $m_F=0$ component.  It can be readily verified that the resulting phases of the components are exactly those of the monopole given in Eq.~(\ref{eq:monopolephase}).  The total time of the laser pulses, assuming that pulses with profiles $\rho_{3}(\mathbf{r})$ and $\rho_{4}(\mathbf{r})$ can be executed simultaneously, is $T=3\pi/\alpha_0$.  Strictly speaking, pulse C is not physically possible --- the profile $\rho_{3}(\mathbf{r})$ requires an axis of propagation $\hat{k}$ in the $x\mathrm{-}y$ plane, which is inconsistent with a polarization vector of $\hat{\epsilon}=-\frac{\hat{x}+i\hat{y}}{\sqrt{2}}$.  However, it is possible to create an effective interaction term resembling $\hat{\epsilon}\cdot\mathbf{r}$ using a beam that does propagate in the $x\mathrm{-}y$ plane.  Indeed, consider a pulse with polarization vector $\hat{\epsilon}^{\prime}=-\frac{\hat{z}+i\hat{y}}{\sqrt{2}}$ sandwiched between magnetic field pulses that generate the rotations $\mathrm{exp}\left(-i\frac{\pi}{2}F_y\right)$ and $\mathrm{exp}\left(i\frac{\pi}{2}F_y\right)$.  One can readily check that
\be
\mathrm{exp}\left(-i\frac{\pi}{2}F_y\right)\: \hat{\epsilon}^{\prime}\cdot\mathbf{r}\: \mathrm{exp}\left(i\frac{\pi}{2}F_y\right)=\hat{\epsilon}\cdot\mathbf{r}.
\ee
Physically, the $m_F$ eigenstates along the $z$-axis are transformed by rotation to become $m_F$ eigenstates along the $x$-axis.  The phase shifts are subsequently performed on these states, where $\rho_{3}(\mathbf{r})$ and $\hat{\epsilon}^{\prime}$ are consistent, before the system is rotated back.
\begin{table*}[t]
\begin{center}
\begin{tabular}{|c|c|c|c|c|}
\hline
Pulse & Polarization & Intensity profile & Time (in units of $\pi/\alpha_0)$ \\
\hline
A & $\sigma^+$ & $\rho_{1}(\mathbf{r})$ & $1$ \\
B & $\sigma^-$ & $\rho_{2}(\mathbf{r})$ & $1$ \\
C & $\sigma^+$ & $\rho_{3}(\mathbf{r})$ & $1$ \\
D & $\pi$ & $\rho_{4}(\mathbf{r})$ & $1/2$ \\
\hline
\end{tabular}
\caption{The pulse sequence to create the monopole is shown, specifying the beam polarizations, intensity profiles, and pulse durations.  Strictly speaking, the polarization and intensity profile of pulse C are inconsistent, but one can achieve the desired effect using a laser pulse inserted between two magnetic field pulses.\label{table:pulsesequence}}
\end{center}
\end{table*}

Experimentally, one should seek to make $\delta$ as large as possible while increasing $I_0$ to maintain a given potential depth.  This is because the photon scattering rate $\Gamma_{sc}$ of an individual atom scales as $I_0(\Gamma/\delta)^2$ compared to $I_0\Gamma/\delta$ for the dipole potential~\cite{grimm00}, and also because the approximation errors in Eqs.~(\ref{eq:pi-phaseshift})-(\ref{eq:minus-phaseshift}) scale as $\Gamma/\delta$.  To maintain the validity of Eqs.~(\ref{eq:pi-phaseshift})-(\ref{eq:minus-phaseshift}), however, one must impose the limit $\delta\ll{\Delta}E_{\mathrm{fs}}$, where ${\Delta}E_{\mathrm{fs}}\; {\approx}\; 3{\times}10^{12}\ \mathrm{s^{-1}}$ is the fine structure splitting between ${}^{2}P_{1/2}$ and ${}^{2}P_{3/2}$.  Physically, this means that the light shifts due to the excited ${}^{2}P_{3/2}$ states remain negligible compared to those originating from the ${}^{2}P_{1/2}$ states.  As a concrete example, we consider the case $\delta=3{\times}10^{10}\ \mathrm{s^{-1}}$ and $I_0=10^{2}I_s$, where $I_s$ is the saturation intensity.  This results in a pulse sequence length of $T\; {\approx}\; 35\ \mu\mathrm{s}$ and a scattering time of an individual atom of $T_{sc}\; {\equiv}\; 1/\Gamma_{sc}\; {\approx}\; 25T$.  We also mention that the ratio $T_{sc}/T$ can be enhanced by pairing every beam in the original sequence with an appropriately chosen second beam.  If both beams have the same polarization and are detuned equally above and below the ${}^{2}S_{1/2}{\rightarrow}{}^{2}P_{3/2}$ transition, the light shifts from the ${}^{2}P_{3/2}$ states are approximately cancelled out, while those from the ${}^{2}P_{1/2}$ states are reinforced.  This makes it possible to abandon the requirement $\delta\ll{\Delta}E_{\mathrm{fs}}$ and choose much larger detunings.

Finally, we note that the solution (\ref{eq:phaseshift}) for the dynamics of the system under the laser beams is only approximate and overlooks subtler but important points.  Clearly such a transformation is unphysical in our case at the origin as it would introduce an infinite phase gradient energy~\cite{dobrek99}, and a more careful analysis of the dynamics near the center would have to include other terms in the Hamiltonian.  Also, in a real system, an absorption plate cannot perfectly generate the intensity profiles given in~(\ref{eq:+azimuthalintensity}) and~(\ref{eq:-azimuthalintensity}), which leads to various degrees of imperfection in vortex creation~\cite{dobrek99}.  In addition, experimental efforts to create a soliton in a scalar condensate using an intensity profile like that of~(\ref{eq:stepintensity}) show that the dipole potential also creates a density wave in addition to the soliton~\cite{burger99,denschlag00}.  We refer the reader to the cited references for further discussion. Ultimately, however, it does not appear that these imperfections preclude the creation of vortices and solitons, which suggests that the creation of monopoles via phase imprinting should also be viable in spite of such imperfections.

%
\section{Summary}
%
We have proposed a method to create a monopole in an antiferromagnetic spin-1 BEC by extending the phase imprinting concept used in scalar BEC's.  We demonstrate that laser beams with different polarizations generate energy shifts that can be used to engineer the phases of the magnetic substates in a broad manner.  We also present an explicit pulse sequence that, starting from the ground state of the condensate, produces phases in the substates corresponding to those of a monopole.

\begin{acknowledgments}
We are grateful to K.-A. Suominen for fruitful discussions and comments on this work.  D.C. is supported by CIMO through a Fulbright Fellowship.
\end{acknowledgments}
%
%

%
%
%

\end{document}